\documentstyle[aaspp4,psbox,11pt]{article}

\received{2001 August 31}
\accepted{2001 October 4}

\slugcomment{accepted for publication in ApJL}

\begin{document}

\newcommand{ \Vec }[1]{ \mbox{\boldmath $ #1 $} }

\title
{Saturation and Thermalization of the Magnetorotational Instability: \\
Recurrent Channel Flows and Reconnections}

\author
{Takayoshi Sano\altaffilmark{1} and Shu-ichiro Inutsuka\altaffilmark{2,3}}

\altaffiltext{1}
{Department of Astronomy, University of Maryland, College Park, MD
20742-2421; sano@astro.umd.edu}
\altaffiltext{2}
{National Astronomical Observatory, Mitaka, Tokyo 181-8588, Japan}
\altaffiltext{3}
{Current address: Department of Physics, Kyoto University, Kyoto
606-8502, Japan}

\begin{abstract}
The nonlinear evolution and the saturation mechanism of the
 magnetorotational instability (MRI) are investigated using
 three-dimensional resistive MHD simulations.
A local shearing box is used for our numerical analysis and the
 simulations begin with a purely vertical magnetic field.
We find that the magnetic stress in the nonlinear stage of the MRI is
 strongly fluctuating.
The time evolution shows the quasi-periodic recurrence of spike-shape
 variations typically for a few orbits which correspond to the rapid
 amplification of the magnetic field by the nonlinear growth of a
 two-channel solution followed by the
 decay through magnetic reconnections.
The increase rate of the total energy in the shearing box system is
 analytically related to the volume-averaged torque in the system. 
We find that at the saturated state this energy gain of the system is
 balanced with the increase of the thermal energy mostly due to the
 joule heating.
The spike-shape time evolution is a general feature of the nonlinear
 evolution of the MRI in the disks threaded by vertical fields 
 and can be seen if the effective magnetic Reynolds number is larger
 than about unity.
\end{abstract}

\keywords{accretion, accretion disks --- diffusion --- instabilities ---
MHD --- turbulence}

\section{INTRODUCTION}

The understanding of the origin of angular momentum transport has been
required for the development of accretion disk theory.
Balbus \& Hawley (1991) have shown that weakly magnetized accretion
disks are subject to a powerful local MHD instability.
\nocite{bh91}
The discovery of this magnetorotational instability (MRI) brought about
huge progress in the theoretical picture of accretion disks.
Numerical simulations have revealed that the nonlinear evolution of the
MRI 
leads to MHD turbulence in which angular momentum is transported by the
Maxwell stress 
(e.g., Hawley, Gammie, \& Balbus 1995).
\nocite{hgb95}
Because the saturation amplitude of the stress determines the structure
and evolution of the disks, the mechanism of the nonlinear saturation
has a great importance for the theory.
However the mechanism which regulates the saturation amplitude of the
MHD turbulence is still unclear.

The purpose of this letter is to examine what is happening at the
saturated turbulent state in order to find a clue to the understanding
of the saturation mechanism. 
A local shearing box approximation is used for our numerical
calculations so the evolution of the instability can be followed for
many orbital periods.
According to previous numerical studies, the saturation amplitudes of
the Maxwell stress take similar values for both local and global disk 
calculations (e.g., Stone et al. 1996).
\nocite{shg96}
Thus we should find the essential processes of the saturation mechanism
in the local shearing box simulations.

The calculations include the effect of ohmic dissipation.
Local 3D simulations show the nonlinear saturation of the MRI for both
with and without the magnetic resistivity (Matsumoto \& Tajima 1995;
Fleming, Stone, \& Hawley 2000).
\nocite{mt95}
\nocite{fsh00}
In axisymmetric 2D ideal MHD simulations with initially uniform vertical
field, the evolution of the MRI results in an exponentially growing
solution (two-channel flow) with no nonlinear saturation (Hawley \&
Balbus 1992).
\nocite{hb92}
When the effect of ohmic dissipation is efficient enough, however, the
nonlinear saturation occurs even in 2D (Sano, Inutsuka, \& Miyama 1998).
\nocite{sim98}
These results suggest that the resistivity is an important quantity in
determining the saturated state. 
In this letter we focus on the effect of ohmic dissipation in 3D
calculations.
This approach 
has an advantage of reducing the influence 
of numerical dissipation on the saturated state that is characterized by
the physical (i.e., ohmic) dissipation process. 

\section{NUMERICAL CALCULATIONS}

The resistive MHD equations are solved by using a finite-difference code
developed by Sano, Inutsuka, \& Miyama (1999).
\nocite{sim99}
We use the local shearing box approximation described in detail by
Hawley et al. (1995).  
\nocite{hgb95}
The Cartesian coordinate ($x$, $y$, $z$) is defined around the fiducial
point comoving with a fiducial angular velocity $\Omega$.
Initially all the physical quantities are spatially uniform ($\rho =
\rho_0$ and $P = P_0$, where $\rho_0$ and $P_0$ are constant) except for
the azimuthal velocity $v_y (x) = - q \Omega x$, where $q = 3 / 2$ for a
Keplerian disk.
In this letter, we focus on an initial magnetic field configuration with
a uniform vertical field, $B_z = B_{z0}$, although the nonlinear
evolution of the MRI are affected by the field geometries (e.g., Fleming
et al. 2000).
\nocite{fsh00}

We choose normalizations with $\rho_{\rm 0} = 1$, $\Omega = 10^{-3}$ and 
the computational box has radial size $L_x = 1$, azimuthal size $L_y =
4$, and vertical size  $L_z = 1$.
Most of the runs use a standard grid resolution of $32 \times 128 \times 
32$.
The initial characteristic wavelength of the MRI is given by
$\lambda_{\rm MRI} = 2 \pi v_{{\rm A}z0} / \Omega$ for the ideal MHD
case (Balbus \& Hawley 1991), where $v_{{\rm A}z0} = B_{z0} / \sqrt{4
\pi \rho_0}$ is the Alfv{\'e}n speed.
Its ratio to the vertical box size is assumed to be $\lambda_{\rm MRI} /
L_z = 0.32$ for all models.
\nocite{bh91}
Then the initial field strength of the models is $v_{{\rm A}z0} = 5.0
\times 10^{-5}$.  
Note that the net flux associated with the initial vertical field
remains unchanged in the time evolution even if the resistivity is
included.
We assume the initial ratio of the gas and magnetic pressure is $\beta_0
\equiv P_0 / (B_{z0}^2 / 8 \pi) = 10^4$, then the initial gas pressure
is $P_0 = 1.25 \times 10^{-5}$ and the sound speed is $c_{s0} = 4.56
\times 10^{-3}$ with $\gamma = 5/3$.
The magnetic diffusivity $\eta$ is assumed to be constant and
characterized by the magnetic Reynolds number $Re_{M}$.
We define the magnetic Reynolds number as $Re_{M} = VL / \eta \equiv
v_{{\rm A}z0}^2 / \eta \Omega$ using typical velocity $V = v_{{\rm
A}z0}$ and typical length scale $L = v_{{\rm A}z0} / \Omega$.


\section{LONG TIME EVOLUTIONS OF THE MRI}

The efficiency of angular momentum transport is given by the turbulent
shear stress, 
$w_{xy} = - {B_x B_y}/{4 \pi} + \rho v_x \delta v_y$,
where the first and second terms are the Maxwell and Reynolds stress,
respectively, and $\delta v_y \equiv v_y + q \Omega x$.
Figure 1 shows the time evolutions of volume-averaged Maxwell stress,
$\langle - B_x B_y / 4 \pi \rangle$.%
\footnote{The single brackets $\langle f \rangle$ imply a volume average 
of quantity $f$. We also use double brackets $\langle \negthinspace
\langle f \rangle \negthinspace \rangle$ to denote a time and volume
average.}
The top panel is the result of the case with $Re_{M} = 1$ and 
the bottom is of the ideal MHD ($\eta = 0$) case.

The Maxwell stress dominates the Reynolds stress throughout the
turbulent stage by a factor of about 6.
The stress of the nonlinear stage is highly fluctuating for both cases
shown in Figure 1.
The main feature of the time evolutions is the many spike-shape
excursions in the stress.
Power spectral density declines with frequency $w$ approximately as
$w^{-2}$ at frequencies above a few orbital period, and almost flat at
lower frequencies.
Note that the break point corresponds to the typical timescale of the
spike.

The time averages of the Maxwell stress are shown in Figure 1 by dotted
lines and they are 0.024 $P_0$ and 0.013 $P_0$ for the $Re_{M} = 1$ and
$\eta = 0$ runs.
Because the contribution of the spike-shape variations to the average is
large, the understanding of the time evolution of these spikes is
essential to estimate the saturation amplitude of the stress.

The diffusion length $l_{\rm diff} \equiv \eta / v_{\rm Az0} = 0.05$ is
larger than the grid scale $l_{\rm grid} = 1/32 = 0.03$ for the $Re_{M}
= 1$ run, so that we can resolve the physical effect of the
resistivity. 
In the ideal MHD run, on the other hand, the dissipation of the magnetic
field is due to artificial numerical effects (Hawley et al. 1995). 
Although the resemblance of the time evolutions between the $Re_{M} = 1$
and $\eta = 0$ runs might suggest that the numerical dissipation works
in a similar way to the physical resistivity, we use the results of the
$Re_{M} = 1$ run in the following discussions.

At the peak of each spike, 
a well-organized two-channel solution dominates in the entire 
computational box.
Figure 2 shows images of the azimuthal component of the magnetic field,
that is the dominant component at the saturated state.
Top panel is the image at a time of the peak of a spike.
The distribution is almost independent of $x$ and $y$.
The upper half of fluid has a strong horizontal field with positive
$B_x$ and negative $B_y$, and the lower half has the oppositely directed
field with negative $B_x$ and positive $B_y$.
The perturbed velocity is nearly horizontal in the same direction as the
fields but has the maximum at the neutral sheet of the fields.
Because the gas pressure is still larger than the magnetic pressure
by more than two orders of magnitude even at the nonlinear stage, the
evolution is almost incompressible.
Even at the maximum growth of the channel flow, the fluctuations of the
density and pressure are very small, $\langle \delta \rho^2
\rangle^{1/2} / \langle \rho \rangle \sim 7.2 \times 10^{-3}$ and
$\langle \delta P^2 \rangle^{1/2} / \langle P \rangle \sim 1.5 \times
10^{-2}$, compared with the fluctuation of the magnetic pressure,
$\langle \delta P_{\rm mag}^2 \rangle^{1/2} / \langle P_{\rm mag}
\rangle \sim 0.75$. 
The exponentially growing channel solution is an exact solution for
nonlinear incompressible MHD equations, thus the horizontal fields are
amplified up to $|B_{x}|_{\max} \sim 18 B_{z0}$ and $|B_{y}|_{\max} \sim
30 B_{z0}$ by this nonlinear growth. 
The perturbed velocity is of the order of the Alfv{\'e}n speed at this
time and is still subsonic; $\langle | \delta v | \rangle  \sim 0.56
\langle v_{\rm A} \rangle \sim 0.073 \langle c_{s} \rangle$.

The image at a time just after the peak is shown in the middle panel of
Figure 2.
The channel flow is known to be unstable for the parasitic instability
(Goodman \& Xu 1994).
\nocite{gx94}
Because the growth rate is proportional to the amplitude of the channel
solution, this instability appears after the sufficient growth of the
two-channel flow.
The unstable modes have finite $k_x$ and $k_y$ and their wavelength $2
\pi / (k_x^2 + k_y^2)^{1/2}$ must be longer than the vertical length of
the channel $L_z$.
This kind of pattern occurs in the image of the spatial distribution of
the magnetic field.
The parasitic instability generates vertical motions and gives a chance
for oppositely directed magnetic fields to approach each other.
Then the dissipation of the magnetic field takes place efficiently
through the magnetic reconnections.
Finally the channel flow disappears and a disorganized MHD turbulence
lasts until the next channel flow starts growing (bottom panel of
Fig. 2).
The magnetic energy is an order of magnitude smaller than the peak value 
and is almost equi-partitioned with the perturbed kinetic energy;
$\langle |\delta v| \rangle \sim 0.83 \langle v_{\rm A} \rangle \sim
0.053 \langle c_{s} \rangle$.

\section{ENERGY BUDGET IN THE MHD TURBULENCE}

According to Hawley et al. (1995), we define the total energy within the
shearing box as 
\begin{equation}
\Gamma \equiv 
\int d^3 x \left[ 
\rho \left( \frac{v^2}{2} + \epsilon + \phi \right) + \frac{B^2}{8\pi} 
\right] \;,  
\label{eq:Etotal}
\end{equation}
where $\epsilon$ means the specific internal energy and $\phi = - q
\Omega^2 x^2$ is the tidal expansion of the effective potential. 
Using the evolution equation for the resistive MHD system, the
time-derivative of the above equation gives  
\begin{equation}
         \frac{d \Gamma }{dt}  
=
           q \Omega L_x
           \int_{X} dA 
                          \left(   \rho v_x \delta v_y
                                 - \frac{ B_x B_y }{4\pi}
                          \right) 
~, \label{eq:dEtdt}
\end{equation}
where 
$dA$ is the surface element and the integral is taken over either of the
radial boundary.
Thus the increase rate of the total energy is proportional to the stress
$w_{xy}$ at the radial boundary. 
Balbus \& Papaloizou (1999) have shown that the similar relation in the
cylindrical coordinates holds for the global disk problem.  
\nocite{bp99}
Although the final expression of equation (\ref{eq:dEtdt}) does not
explicitly depend on the amount of resistivity, the joule heating is
crucial for the transformation of the magnetic energy into the thermal
energy (see below). 


Using the volume-averaged stress $\langle w_{xy} \rangle$ instead of the
surface-averaged value of the stress at the radial boundary $\langle
w_{xy} \rangle_{X}$, the volume average of the input energy $\langle
\dot{E}_{\rm in} \rangle$ is approximately given by
\begin{equation}
       \langle \dot{E}_{\rm in} \rangle \equiv 
       \frac{\partial \langle E_{\rm in} \rangle}{\partial t} 
       =   q \Omega \left\langle w_{xy} \right\rangle_X 
       \approx  
           q \Omega \left\langle w_{xy} \right\rangle \;.
\end{equation} 
Because our numerical scheme solves the energy equation in terms of the
total energy, our calculations perfectly satisfy equation
(\ref{eq:dEtdt}). 
Furthermore, the density does not vary much so that the change in
potential energy $\rho \phi$ is negligible compared with the other terms
in equation (\ref{eq:Etotal}).    
Thus the input energy $E_{\rm in}$ is almost identical to the increase
of the sum of $E_{\rm th} = \rho \epsilon = P / (\gamma - 1)$, $E_{\rm
mag} = B^2 / 8 \pi$, and $E_{\rm kin} = \rho v^2 / 2$.
This is clearly shown in Figure 3 which depicts the time evolutions of 
$\langle \dot{E}_{\rm in} \rangle$ and $\langle \dot{E}_{\rm tot}
\rangle \equiv \langle \dot{E}_{\rm th} \rangle + \langle \dot{E}_{\rm
mag} \rangle + \langle \dot{E}_{\rm kin} \rangle$.
$\langle \dot{E}_{\rm in} \rangle$ and $\langle \dot{E}_{\rm tot}
\rangle$ take such close values that their curves overlap each other.

The time evolution of $\langle \dot{E}_{\rm th} \rangle$ is shown in
Figure 3 by a dotted curve.
Because $\langle \dot{E}_{\rm in} \rangle$ is proportional to the
Maxwell stress, its time evolution has spike-shape variations.
As seen from the figure, the evolution of $\langle \dot{E}_{\rm th}
\rangle$ also has many spikes with the similar amplitude to $\langle
\dot{E}_{\rm in} \rangle$.
The time average of $\langle \dot{E}_{\rm th} \rangle$ takes almost the 
same value as that of $\langle \dot{E}_{\rm in} \rangle$.
The difference $|\langle \negthinspace \langle \dot{E}_{\rm th} \rangle
\negthinspace \rangle - \langle \negthinspace \langle \dot{E}_{\rm
in} \rangle \negthinspace \rangle| / \langle \negthinspace \langle
\dot{E}_{\rm in} \rangle \negthinspace \rangle$ is less than $10^{-2}$,
when we calculate the time averages from 10 to 25 orbits.
Thus the energy gain of the system is finally transformed into the
thermal energy.
The thermal energy must be increasing throughout the nonlinear evolution
unless some cooling process is included.
The magnetic and kinetic energies, on the other hand, are saturated,
thus the time averages of their rates of change are $\langle
\negthinspace \langle \dot{E}_{\rm mag} \rangle \negthinspace \rangle
\sim \langle \negthinspace \langle \dot{E}_{\rm kin} \rangle
\negthinspace \rangle \sim 0$.
This means that the time and spatial average of the energy gain $d
\Gamma / dt$ should amount to the time and spatial average of the
heating rate due to turbulent dissipation, provided the magnetic and
kinetic energy are saturated.  
That is 
\begin{equation}
       \langle \negthinspace \langle 
               \dot{E}_{\rm th} 
       \rangle \negthinspace \rangle 
     \approx
         q \Omega
       \langle \negthinspace \langle 
               w_{xy}
       \rangle \negthinspace \rangle \;.
\end{equation}
Here we cannot remove the symbol $\langle \negthinspace \langle \rangle
\negthinspace \rangle$ 
from this equation. 
In other words, this {\em fluctuation-dissipation relation} holds only
in the coarse-grained description of the thermalization rate and the
correlated fluctuation of magnetic field and velocity perturbations. 
Note that this equation does not hold in the 2D simulation with $Re_{M}
\gtrsim 1$ where the system does not show the saturation. 

Figure 4 shows the time evolutions of $\langle \dot{E}_{\rm in} \rangle$,
$\langle \dot{E}_{\rm th} \rangle$, and $\langle \dot{E}_{\rm mag}
\rangle$ at a typical spike-shape variation.
The joule heating rate $\langle \dot{E}_{\rm joule} \rangle \equiv \eta
| \nabla \times \mbox{\boldmath $B$} |^2 / 4 \pi$ is also shown in this
figure.
The joule heating is about 80 percent of the thermal energy increase. 
(This fraction is almost 100 \% for the $Re_{M}=0.3$ run.)
Throughout the nonlinear evolution, the magnetic energy is transformed
into the thermal energy at a rate which over time matches $\langle
\dot{E}_{\rm in} \rangle$.
Thus the saturation of the magnetic energy is achieved by a
time-averaged balance between the energy gain by the growth of the MRI
and the loss by the joule heating in the reconnection process.

We find that the peak in $\langle \dot{E}_{\rm th} \rangle$ is always
delayed from $\langle \dot{E}_{\rm in} \rangle$, and that the sign of
$\langle \dot{E}_{\rm mag} \rangle$ changes around the peak in $\langle
\dot{E}_{\rm th} \rangle$.
The kinetic energy has a similar time evolution to
$E_{\rm mag}$, but the amplitude of the fluctuations in $\langle
\dot{E}_{\rm kin} \rangle$ is negligible compared with that of the
magnetic energy.
The spike-shape variation of $\langle \dot{E}_{\rm in} \rangle$ starts
with the onset of a channel flow.
The magnetic energy increases due to the growth of the MRI and 
$\langle \dot{E}_{\rm mag} \rangle$ takes a positive peak value
at the maximum growth of the two-channel solution, which corresponds to
the peak in $\langle \dot{E}_{\rm in} \rangle$ (top panel of Fig. 2).
When the channel flow is destroyed by the parasitic instability,
the large scale magnetic reconnections become efficient.
Then the sign of $\langle \dot{E}_{\rm mag} \rangle$ changes from
positive to negative and the thermal energy increases rapidly.
When the $\langle \dot{E}_{\rm mag} \rangle$ takes a negative peak,
$\langle \dot{E}_{\rm th} \rangle$ is nearly at the top of the peak
(middle panel of Fig. 2).

\section{DEPENDENCE OF THE SATURATION CHARACTER ON THE MAGNETIC REYNOLDS
 NUMBER}

We find that two-channel solutions appear repeatedly in nonlinear turbulent
stage. 
The behavior similar to this appearance of a channel flow can be seen in
2D simulations, which show a two-channel flow appears at the end of the
calculations even though the initial value of $\lambda_{\rm MRI}$ is
smaller than the box size.
This inverse cascade is characteristic of both 2D and 3D nonlinear
evolutions of the MRI in the cases $Re_{M} \gtrsim 1$.
The difference between 2D and 3D simulations is that in 3D the channel
flow is unstable to non-axisymmetric modes of the parasitic instability.
Then the question is why the two-channel solution appear in nonlinear
stage of the MRI.

The key quantity is the characteristic length of the instability
expected from the linear analysis.
The most unstable wavelength of the MRI is given by $\lambda_{\rm MRI}
\sim v_{\rm A} / \Omega \propto v_{\rm A}$ for less resistive cases
$Re_{M} \gtrsim 1$.
If the field is amplified by the growth of the MRI, then $\lambda_{\rm
MRI}$ shifts to the longer wavelength.
Therefore it can reach the vertical box size, and the longest wavelength
mode, a two-channel solution, appears finally in nonlinear evolution. 
This is the reason for the inverse cascade.
For $Re_{M} \lesssim 1$, on the other hand, the characteristic
wavelength is given by $\lambda_{\rm MRI} \sim \eta / v_{\rm A} \propto
v_{\rm A}^{-1}$ (Sano \& Miyama 1999). 
\nocite{sm99}
Thus the wavelength $\lambda_{\rm MRI}$ becomes shorter when the field
is amplified.
In fact, 2D simulations with $Re_{M} \lesssim 1$ show that the saturated
state is a disorganized MHD turbulence and no emergence of a two-channel
flow (Sano et al. 1998).
\nocite{sim98}
We find this is true even for 3D cases.
For the case with $Re_{M} = 0.3$, the nonlinear evolution shows no
growth of a two-channel flow.
The amplitude of the time variability of the Maxwell stress is very
small in this case.
Due to the lack of the nonlinear growth of the channel flow, 
the horizontal component of the magnetic energy is not amplified 
very much and remains 
only 2 times larger than the vertical, while this ratio $\langle
\negthinspace \langle B_y^2 \rangle \negthinspace \rangle / \langle
\negthinspace \langle B_z^2 \rangle \negthinspace \rangle$ is an order of
magnitude larger for $Re_{M} \gtrsim 1$ cases.
The energy balance at the saturated state is roughly given 
by $\langle \negthinspace \langle \dot{E}_{\rm in} \rangle \negthinspace
\rangle \sim \langle \negthinspace \langle \dot{E}_{\rm th} \rangle
\negthinspace \rangle \sim \langle \negthinspace \langle \dot{E}_{\rm
joule} \rangle \negthinspace \rangle$ same as in the cases with $Re_{M}
\gtrsim 1$.
Because the diffusion time is comparable to the growth time of the MRI
in this case, the magnetic diffusion would be an important process for
the joule heating as well as the recurrent and rapid magnetic
reconnections.
The saturation amplitude is two orders of magnitude smaller than the
models with $Re_{M} \gtrsim 1$, so that the angular momentum transport
by the MRI is not efficient when $Re_{M} \lesssim 1$.

We can categorize the saturated states of the MRI into two types when
the disks are threaded by uniform vertical fields initially.
The first type involves the recurrence of the channel flow and large
scale reconnections (the $Re_{M} = 1$ and ideal MHD runs). 
In the disorganized MHD turbulence of the second type, the magnetic
diffusion and reconnections prevail without the significant growth of
the channel flow (the $Re_{M} = 0.3$ run).
The angular momentum transport by the magnetic stress is efficient only
for the first type. 
The difference of these two types is caused by the different
dependence of the characteristic length of the MRI, 
$\lambda_{\rm MRI}$, on the magnetic field strength.
Thus the magnetic Reynolds number defined as $Re_{M} = v_{\rm A}^2 /
\eta \Omega$ can characterize the linear and nonlinear features
of the MRI and the critical value is $Re_{M} \sim 1$.


\acknowledgements
We thank James Stone and Neal Turner for useful discussions.
Computations were carried out on the VPP5000 at the National
Astronomical Observatory of Japan and the VPP700 at the Subaru
Telescope, NAOJ.



\clearpage

\begin{figure}
\begin{center}
\hspace*{-45pt}
\setlength{\unitlength}{0.1bp}%
{
}%
\begin{picture}(3600,1278)(0,0)%
{
}%
\put(1242,1039){\makebox(0,0)[l]{\large{$Re_{M} = 1$}}}%
\put(2250,50){\makebox(0,0){ }}%
\put(700,739){%
\makebox(0,0)[b]{\shortstack{\large{\hspace{-90pt} $\langle - B_x B_y / 4 \pi \rangle / P_0$}}}%
}%
\put(1000,993){\makebox(0,0)[r]{\large $0.15$}}%
\put(1000,762){\makebox(0,0)[r]{\large $0.1$}}%
\put(1000,531){\makebox(0,0)[r]{\large $0.05$}}%
\put(1000,300){\makebox(0,0)[r]{\large $0$}}%
\end{picture}%
\\ \vspace{-41pt}
\hspace*{-45pt}
\setlength{\unitlength}{0.1bp}%
{
}%
\begin{picture}(3600,1278)(0,0)%
{
}%
\put(1242,1039){\makebox(0,0)[l]{\large{$\eta = 0$}}}%
\put(2250,50){\makebox(0,0){\large $t / t_{\rm rot}$}}%
\put(700,739){%
\makebox(0,0)[b]{\shortstack{ }}%
}%
\put(3450,200){\makebox(0,0)[c]{\large $100$}}%
\put(2850,200){\makebox(0,0)[c]{\large $75$}}%
\put(2250,200){\makebox(0,0)[c]{\large $50$}}%
\put(1650,200){\makebox(0,0)[c]{\large $25$}}%
\put(1050,200){\makebox(0,0)[c]{\large $0$}}%
\put(1000,993){\makebox(0,0)[r]{\large $0.15$}}%
\put(1000,762){\makebox(0,0)[r]{\large $0.1$}}%
\put(1000,531){\makebox(0,0)[r]{\large $0.05$}}%
\put(1000,300){\makebox(0,0)[r]{\large $0$}}%
\end{picture}%
\\
\caption
{Time evolution of volume-averaged Maxwell stress for the model with
$Re_{M} = 1$ ($top$) and $\eta = 0$ ($bottom$).
Time is given in the rotation time $t_{\rm rot} = 2 \pi / \Omega$.
Dotted lines show their time-averaged values done from 10 to 100 orbits.
The grid resolution of these runs is $32 \times 128 \times 32$.
\label{fig:tm-t}}
\end{center}
\end{figure}
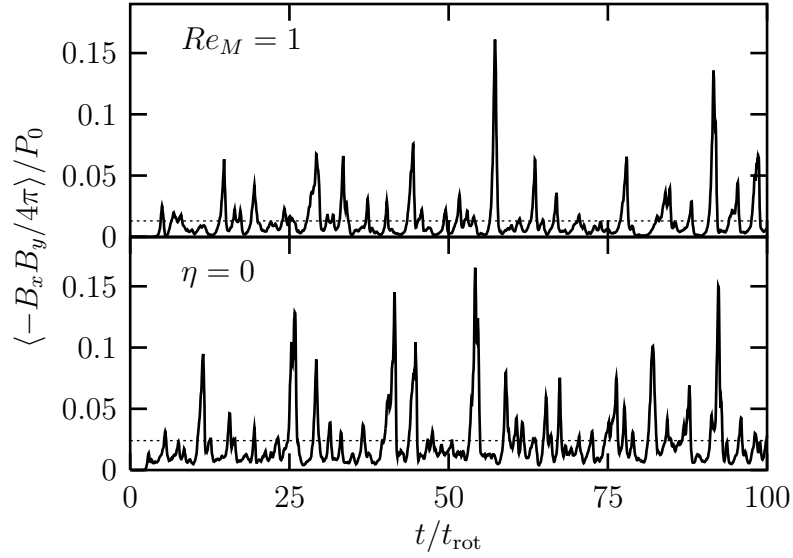
 
\begin{figure}
\begin{center}
\hspace*{-45pt}
\psbox{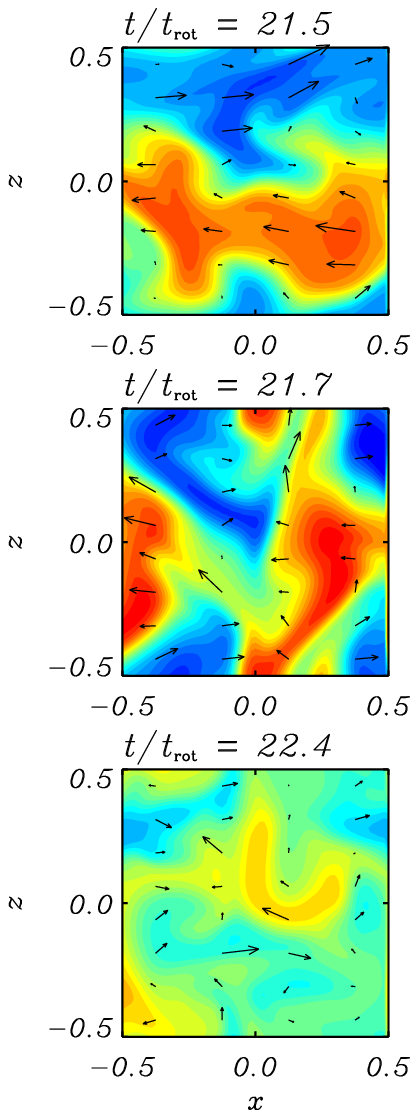}
\hspace*{-70pt}
\psbox{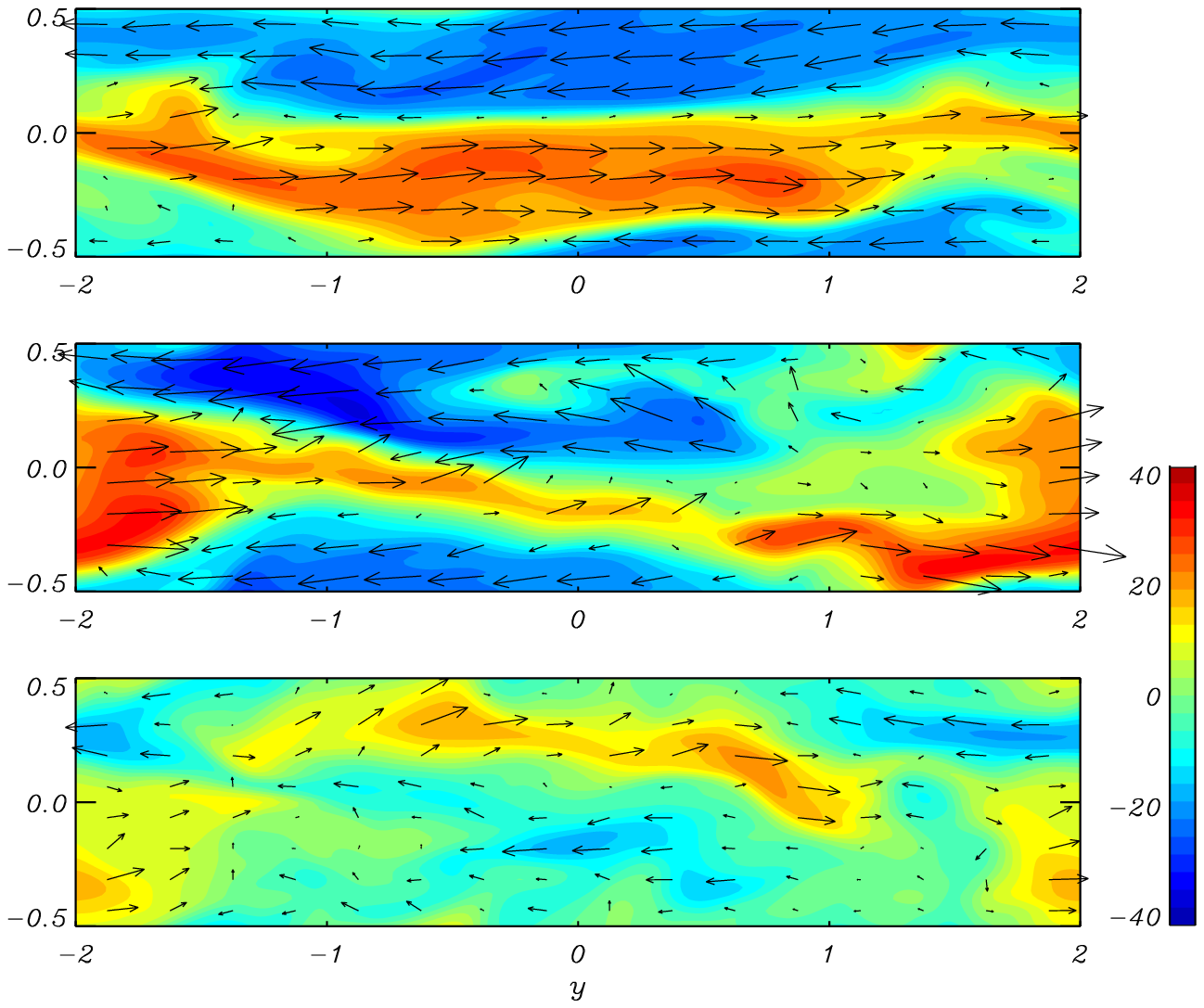}
\caption
{Slices in the $x$-$z$ plane at $y = - 2$ and in the $y$-$z$ plane at $x = 
-0.5$ of the azimuthal component of the magnetic fields $B_y / B_{z0}$
($colors$) and magnetic field vectors ($arrows$) in the $Re_{M} = 1$ run
with grid resolution $64 \times 256 \times 64$.
From top to bottom the time of the image is $t/t_{\rm rot} = 21.5$,
21.7, and 22.4.
\label{fig:imgs}}
\end{center}
\end{figure}

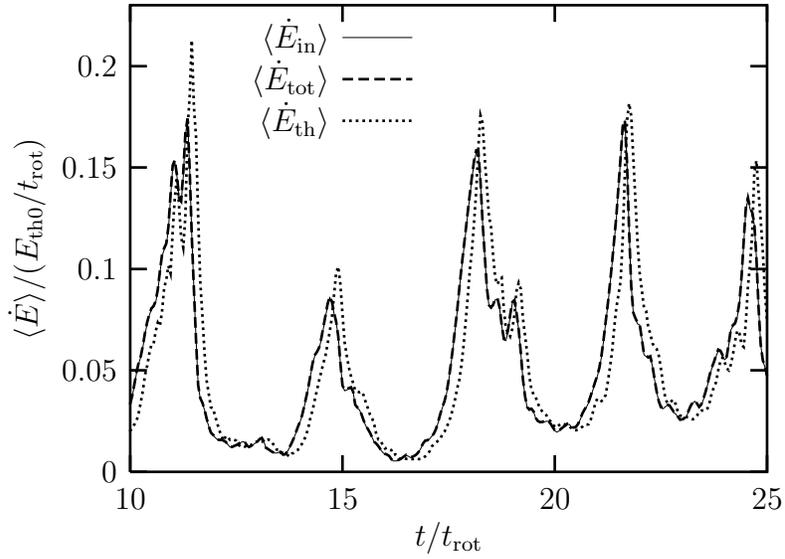
\begin{figure}
\begin{center}
\hspace*{-45pt}
\setlength{\unitlength}{0.1bp}%
{
}%
\begin{picture}(3600,2160)(0,0)%
{
}%
\put(1800,1624){\makebox(0,0)[r]{\large{$\langle \dot{E}_{\rm th} \rangle$}}}%
\put(1800,1781){\makebox(0,0)[r]{\large{$\langle \dot{E}_{\rm tot} \rangle$}}}%
\put(1800,1938){\makebox(0,0)[r]{\large{$\langle \dot{E}_{\rm in} \rangle$}}}%
\put(2250,50){\makebox(0,0){\large $t / t_{\rm rot}$}}%
\put(700,1180){%
\makebox(0,0)[b]{\shortstack{\large{$\langle \dot{E} \rangle / (E_{\rm th0} / t_{\rm rot})$}}}%
}%
\put(3450,200){\makebox(0,0)[c]{\large $25$}}%
\put(2650,200){\makebox(0,0)[c]{\large $20$}}%
\put(1850,200){\makebox(0,0)[c]{\large $15$}}%
\put(1050,200){\makebox(0,0)[c]{\large $10$}}%
\put(1000,1830){\makebox(0,0)[r]{\large $0.2$}}%
\put(1000,1448){\makebox(0,0)[r]{\large $0.15$}}%
\put(1000,1065){\makebox(0,0)[r]{\large $0.1$}}%
\put(1000,683){\makebox(0,0)[r]{\large $0.05$}}%
\put(1000,300){\makebox(0,0)[r]{\large $0$}}%
\end{picture}%
\\
\caption
{Time evolution of volume-averaged time derivative of input energy $\langle
\dot{E}_{\rm in} \rangle$, total energy $\langle \dot{E}_{\rm tot}
\rangle \equiv \langle \dot{E}_{\rm th} \rangle + \langle \dot{E}_{\rm
mag} \rangle + \langle \dot{E}_{\rm kin} \rangle$, and thermal energy
$\langle \dot{E}_{\rm th} \rangle$ of the $Re_{M} = 1$ run with grid
resolution $64 \times 256 \times 64$.
The time derivatives of the energies are given in units of $E_{\rm
th0}/t_{\rm rot}$ where $E_{\rm th0} = P_0 / (\gamma - 1)$ is the
initial thermal energy.
\label{fig:edot}}
\end{center}
\end{figure}

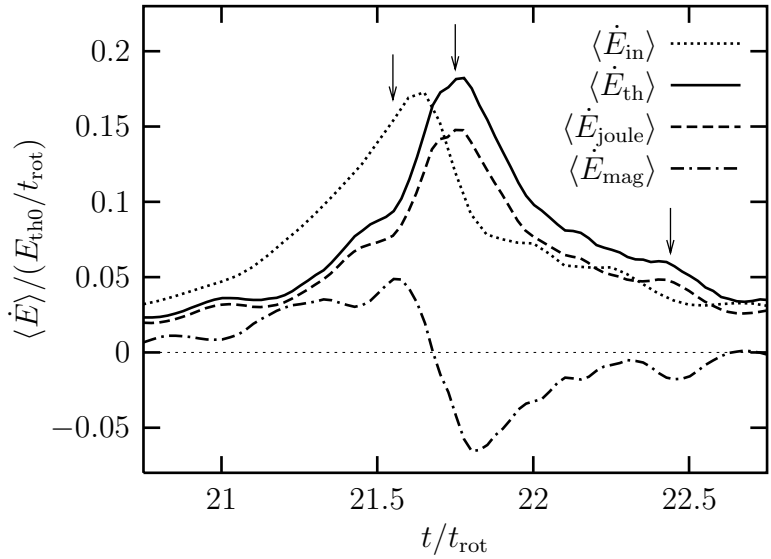
\begin{figure}
\begin{center}
\hspace*{-45pt}
\setlength{\unitlength}{0.1bp}%
{
}%
\begin{picture}(3600,2160)(0,0)%
{
}%
\put(3037,1448){\makebox(0,0)[r]{\large{$\langle \dot{E}_{\rm mag} \rangle$}}}%
\put(3037,1605){\makebox(0,0)[r]{\large{$\langle \dot{E}_{\rm joule} \rangle$}}}%
\put(3037,1762){\makebox(0,0)[r]{\large{$\langle \dot{E}_{\rm th} \rangle$}}}%
\put(3037,1919){\makebox(0,0)[r]{\large{$\langle \dot{E}_{\rm in} \rangle$}}}%
\put(2275,50){\makebox(0,0){\large $t / t_{\rm rot}$}}%
\put(700,1180){%
\makebox(0,0)[b]{\shortstack{\large{$\langle \dot{E} \rangle / (E_{\rm th0} / t_{\rm rot})$}}}%
}%
\put(3156,200){\makebox(0,0)[c]{\large $22.5$}}%
\put(2569,200){\makebox(0,0)[c]{\large $22$}}%
\put(1981,200){\makebox(0,0)[c]{\large $21.5$}}%
\put(1394,200){\makebox(0,0)[c]{\large $21$}}%
\put(1050,1890){\makebox(0,0)[r]{\large $0.2$}}%
\put(1050,1606){\makebox(0,0)[r]{\large $0.15$}}%
\put(1050,1322){\makebox(0,0)[r]{\large $0.1$}}%
\put(1050,1038){\makebox(0,0)[r]{\large $0.05$}}%
\put(1050,754){\makebox(0,0)[r]{\large $0$}}%
\put(1050,470){\makebox(0,0)[r]{\large $-0.05$}}%
\end{picture}%
\\
\caption
{Time evolution of volume-averaged time derivative of input energy
$\langle \dot{E}_{\rm in} \rangle$, thermal energy $\langle \dot{E}_{\rm
th} \rangle$, and magnetic energy $\langle \dot{E}_{\rm mag} \rangle$
and volume-averaged joule heating rate $\langle \dot{E}_{\rm joule}
\rangle$ at a spike-shape variation in $Re_{M} = 1$ run with grid
resolution $64 \times 256 \times 64$.
Arrows denote the times of the snapshots shown in Figure 2.
\label{fig:spike}}
\end{center}
\end{figure}

\end{document}